\documentclass[aps,prl,twocolumn,superscriptaddress]{revtex4-2}
\usepackage{float}
\usepackage{amsmath}
\usepackage{mathtools} 
\usepackage{amssymb}
\usepackage{amsthm}
\usepackage{bm} 
\usepackage{upgreek} 
\usepackage{xcolor}
\usepackage{graphicx}
\usepackage{subfigure}
\usepackage{epstopdf}
\usepackage{ulem}
\usepackage[colorlinks,linkcolor=blue,anchorcolor=blue,citecolor=blue,urlcolor=blue]{hyperref}
\usepackage{braket}

\newcommand{\nn}{{\nonumber}}

\newcommand{\up}{\uparrow}
\newcommand{\dn}{\downarrow}

\newcommand{\av}[1]{\left\langle #1 \right\rangle}

\begin{document}
\title{Effective model and $s_\pm$-wave superconductivity in trilayer nickelate La$_4$Ni$_3$O$_{10}$}

\author{Qing-Geng Yang} \thanks{Q.G.Y. and K.Y.J contributed equally to this work.}
\affiliation{National Laboratory of Solid State Microstructures $\&$ School of Physics, Nanjing University, Nanjing 210093, China}

\author{Kai-Yue Jiang} \thanks{Q.G.Y. and K.Y.J contributed equally to this work.}
\affiliation{School of Physics and Physical Engineering, Qufu Normal University, Qufu 273165, China}

\author{Da Wang} \email{dawang@nju.edu.cn}
\affiliation{National Laboratory of Solid State Microstructures $\&$ School of Physics, Nanjing University, Nanjing 210093, China}
\affiliation{Collaborative Innovation Center of Advanced Microstructures, Nanjing University, Nanjing 210093, China}

\author{Hong-Yan Lu} \email{hylu@qfnu.edu.cn}
\affiliation{School of Physics and Physical Engineering, Qufu Normal University, Qufu 273165, China}

\author{Qiang-Hua Wang} \email{qhwang@nju.edu.cn}
\affiliation{National Laboratory of Solid State Microstructures $\&$ School of Physics, Nanjing University, Nanjing 210093, China}
\affiliation{Collaborative Innovation Center of Advanced Microstructures, Nanjing University, Nanjing 210093, China}

\begin{abstract}
The recent discovery of bulk superconductivity in trilayer nickelate La$_4$Ni$_3$O$_{10}$ with the critical temperature $T_c$ near $30$~K under high pressure is attracting a new wave of research interest, after the breakthrough of bilayer La$_3$Ni$_2$O$_7$ with $T_c$ near $80$~K.
The similarities and differences of electronic structure and  superconducting mechanism in these two systems are urgent theoretical issues.
In this Letter, we study the electronic band structure and construct a minimal trilayer tight-binding model for the high-pressure phase of La$_4$Ni$_3$O$_{10}$ in terms of the nickel $3d_{x^2-y^2}$ and $3d_{3z^2-r^2}$ orbitals, and study the superconducting mechanism due to local Coulomb interactions by the unbiased functional renormalization group. We find antiferromagnetic correlations between the outer layers instead of neighboring ones, apart from the inplane correlations.
The effective interaction induces Cooper pairing with the $s_\pm$-wave symmetry, which changes sign across the Fermi pockets. We find $T_c$ in La$_4$Ni$_3$O$_{10}$ is systematically lower than that in La$_3$Ni$_2$O$_7$, and electron doping can enhance $T_c$.
\end{abstract}
\maketitle

{\it Introduction}.
As an important progress in pursuing high-temperature superconductivity with perovskite structure other than cuprates, the discovery of bilayer Ruddlesden-Popper (RP) La$_3$Ni$_2$O$_7$ (La327) with critical temperature $T_c$ near $80$~K under high pressure \cite{2023Nature} has attracted significant attention both experimentally \cite{exp-9,exp-13,exp-16,exp-28,exp-30,exp-31,exp-32,exp-33,exp-34,exp-35,exp-37,exp-38,exp-41,exp-43,exp-44,exp-48,exp-49,exp-50,exp-51,exp-52,exp-53,exp-54,exp-56,exp-57} and theoretically \cite{t-1,t-2,t-3,t-5,t-6,t-7,t-8,t-60,t-10,t-11,t-12,t-14,t-15,t-17,t-19,t-21,t-22,t-4,t-20,t-23,t-36,t-47,t-59,t-24,t-25,t-26,t-27,t-29,t-58,t-39,t-40,t-42,t-45,327arxivhu2,t-46,t-58}.
Since the average electronic configuration of the nickel atom is $3d^{7.5}$, both $3d_{x^2-y^2}$ and $3d_{3z^2-r^2}$ orbitals are active ones near the Fermi level.
In this material, the superconductivity is now generally attributed to the strong vertical inter-layer coupling between the nickel $3d_{3z^2-r^2}$ orbitals, leading to an extended $s$-wave (also called $s_\pm$-wave) pairing \cite{t-2,t-5,t-6,t-14,t-15,t-17,t-21,t-24,t-25,t-26,t-45,327arxivhu2}, although there are also some other proposals \cite{t-4,t-23,t-20,t-59,t-36,t-47,t-55,t-58}.
Such a picture is quite different from cuprates or infinite-layer nickelates, which have the atomic configuration of Cu or Ni near $3d^9$ and hence can be described by the effective single-orbital model \cite{Anderson1987,Zhang-Rice_1988}.

At the present stage, the superconducting volume fraction of La327 samples appears to be very small \cite{exp-41}, asking for further refinement of the samples. In fact, the bilayer RP phase appears to coexist with another alternated single-triple-layer (``1313'') RP phase \cite{exp-48,exp-49,exp-50}.
Interestingly, the trilayer RP La$_4$Ni$_3$O$_{10}$ (La4310) is also reported recently to show signatures of superconductivity \cite{exp-28,t-29,exp-30,exp-31} and zero resistance \cite{exp-32} under high pressure above $43$~GPa. The maximum $T_c$ of La4310 (reaching up to $30$~K) is lower than La327, while its superconducting volume fraction is found to be very high \cite{exp-32}, indicating the superconductivity in La4310 is a bulk property rather than filamentary as in the present-stage La327 \cite{exp-41}.
There are some similarities between the two systems.
La4310 has a nominal chemical valence Ni$^{2.67+}$, in favor of a mixed $3d^7$ and $3d^8$ electronic configuration as in La327.  La4310 also undergoes a structural transition from monoclinic $P21/a$ to tetragonal $I4/mmm$ space group at pressures around $12$-$20$ GPa \cite{t-29,exp-32,exp-34}, together with the change of the $c$-axis Ni-O-Ni bond angle into 180$^\circ$. The superconductivity only occurs in the high pressure tetragonal phase. There is, however, a marked difference between the two families: the La4310 has three layers, and the inner one is unequivalent to the outer ones. It is a timely and important issue to investigate the pairing mechanism and pairing symmetry of La4310, and to understand how it is related to La327.

In this work, we construct a minimal trilayer tight-binding model with the nickel $3d_{3z^2-r^2}$ and $3d_{x^2-y^2}$ orbitals for La4310 in the high-pressure tetragonal phase from {\it ab initio} calculations on the basis of first-principle calculations. We then investigate the electronic correlation effect from two-orbital Coulomb interactions using the unbiased singular-mode functional renormalization group (SM-FRG)\cite{Wang_PRB_2012,Wang_PRB_2013,Tang_PRB_2019,t-2}. We find the spin fluctuations exhibit a peculiar antiferromagnetic coupling between the two outer-layers, which subsequently induces the $s_{\pm}$-wave superconductivity with the dominant pairings between $3d_{3z^2-r^2}$ orbitals both in the inner-plane and between the two outer-planes rather than neighboring ones. Our results show $T_c$ of La4310 is systematically lower than that of La327, setting La327 as the most promising candidate for high-$T_c$ in the RP families of nickelates.

\begin{figure*}
\includegraphics[width=\linewidth]{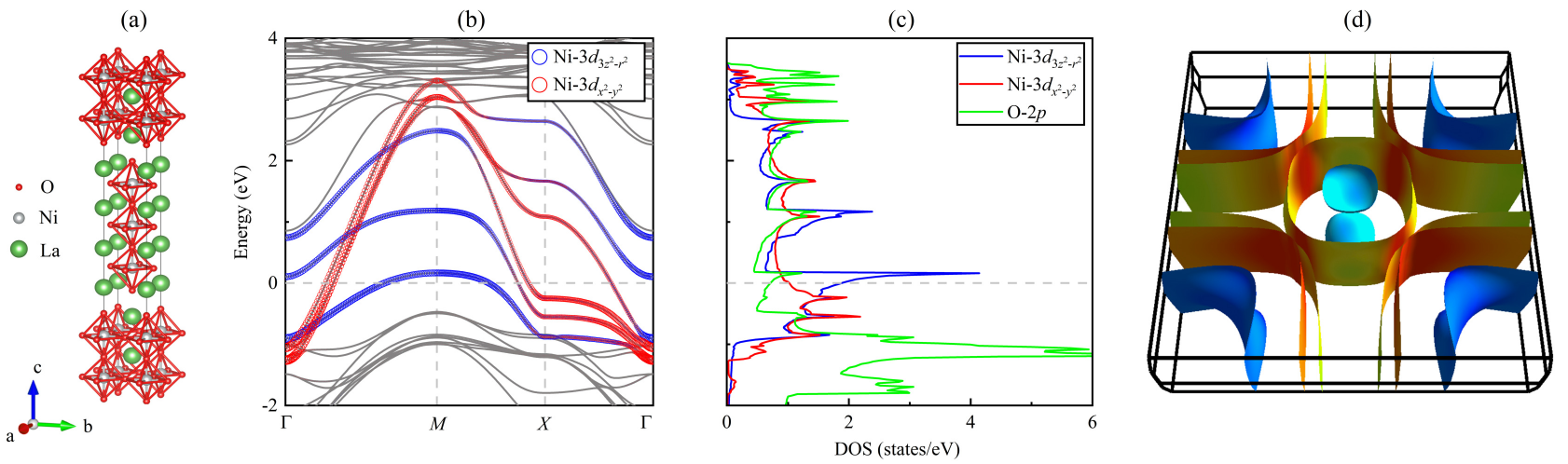}
\caption{(a) Crystal structure, (b) orbital-projected band structure, (c) partial DOS, and (d) three-dimensional Fermi surface of La4310 in the high-pressure tetragonal phase.}
\label{fig:lda}
\end{figure*}

{\it Tight-binding model}.
For the trilayer RP La4310 under high pressure with the tetragonal lattice structure shown in Fig.~\ref{fig:lda}(a), we firstly perform density functional theory (DFT) calculations by Vienna {\it ab initio} simulation package (VASP) \cite{VASP}.
The crystal structure parameters with $I$4/$mmm$ space group at $40$~GPa are adopted \cite{exp-34}. The projector augmented-wave (PAW) \cite{PAW} method and the generalized gradient approximation (GGA) of Perdew-Burke-Ernzerhof (PBE) \cite{PBE} exchange-correlation functional are adopted. The plane-wave cutoff energy is set as $500$~eV. The $\Gamma$-centered $12\times12\times12$ and $36\times36\times36$ $k$-point grids are used for the self-consistent and density of states (DOS) calculations, respectively.
Combining the orbital-projected band structure and partial DOS as shown in Figs.~\ref{fig:lda}(b) and \ref{fig:lda}(c), it is evident that the states near the Fermi level ($E_\mathrm{F}$) are dominated by nickel $3d_{x^2-y^2}$ and $3d_{3z^2-r^2}$ orbitals, which is consistent with the chemical valence analysis and similar to La327 \cite{t-1,t-3,2023Nature,t-5,t-6,327arxivhu2}.
In Fig.~\ref{fig:lda}(d), we show the three-dimensional Fermi surface. The four sheets of large Fermi surfaces are found to be roughly two-dimensional-like, except the additional small Fermi sphere enclosing $(0,0,\pi)$.

\begin{table}[b]
	\caption{Onsite energies $\varepsilon_a$ and hopping integrals $t_\delta^{ab}$ of the trilayer two-orbital tight-binding model for La4310 at 40 GPa. Here, $x$/$z$ denotes the $3d_{x^2-y^2}$/$3d_{3z^2-r^2}$ orbital, I/O denotes the inner/outer-layer. Note that the inter-layer distance is denoted by $\frac12$. The energy is in unit of eV.}
	\centering
	\setlength{\tabcolsep}{5mm}
	\renewcommand{\arraystretch}{2}
	\begin{tabular}{cccc}
		\hline
		\hline
		I-$\varepsilon_x$ & I-$\varepsilon_z$ & O-$\varepsilon_x$ & O-$\varepsilon_z$  \\ \hline
		1.116 & 1.059 & 0.858 & 0.676  \\ \hline \hline
		I-$t_{(100)}^{x x}$ & I-$t_{(100)}^{z z}$ & O-$t_{(100)}^{x x}$ &O-$t_{(100)}^{z z}$  \\ \hline
		-0.507 & -0.158 & -0.507 & -0.146  \\ \hline \hline
		I-$t_{(100)}^{x z}$ & I-$t_{(110)}^{xx}$ & O-$t_{(100)}^{x z}$ & O-$t_{(110)}^{xx}$ \\ \hline
		0.282 & 0.063 & 0.275 & 0.069  \\ \hline \hline
	\end{tabular}
	\centering
	\setlength{\tabcolsep}{4mm}
	\renewcommand{\arraystretch}{2}
	\begin{tabular}{ccccc}
		$t_{(00\frac{1}{2})}^{z z}$ & $t_{(00\frac{1}{2})}^{x x}$ & $t_{(001)}^{z z}$ & $t_{(10\frac{1}{2})}^{x z}$ & $t_{(10\frac{1}{2})}^{z z}$   \\ \hline
		-0.684 & 0.004 & -0.087 & -0.039 & 0.036  \\
		\hline
		\hline
	\end{tabular}
	\label{table:tb}
\end{table}

\begin{figure}[b]
\includegraphics[width=\linewidth]{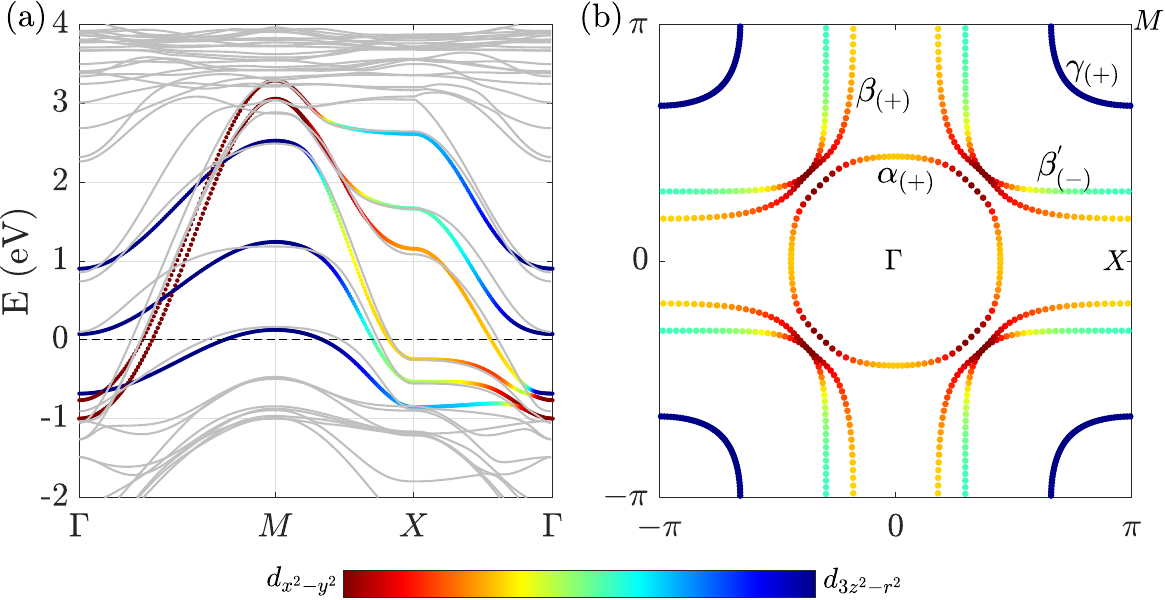}
\caption{(a) Band structure and (b) Fermi surfaces obtained from the tight-binding model. The DFT band at $k_z=0$ (gray lines) is shown for comparison.
The orbital weights are represented by colors and the $z$-reflection parities $P_z$ are indicated as the subscripts of the pocket names.}
\label{fig:tb}
\end{figure}

Subsequently, the maximally localized Wannier functions \cite{Wannier1,Wannier2} with $d_{x^2-y^2}$ and $d_{3z^2-r^2}$ symmetries centering at each Ni-atom are extracted by the state-of-the-art Wannier90 \cite{Wannier3} to construct a trilayer two-orbital tight-binding model
\begin{equation}
H_0=\sum_{i \delta, a b, \sigma} t_\delta^{a b} c_{i a \sigma}^{\dagger} c_{i+\delta b \sigma}+\sum_{i a \sigma} \varepsilon_a c_{i a \sigma}^{\dagger} c_{i a \sigma},
\end{equation}
where $t_\delta^{a b}$ is the hopping matrix element between the $a$ orbital on site $i$ and the $b$ orbital on site $i$ + $\delta$, $\sigma$ denotes spin, and $\varepsilon_a$ is the on-site energy of the $a$-orbital. In the following, we use $x$/$z$ to denote the $3d_{x^2-y^2}$/$3d_{3z^2-r^2}$ orbital for simplicity. Up to $D_{4h}$ symmetry operations, the onsite energy and hopping parameters are listed in Table~\ref{table:tb}.
It is found that the vertical inter-layer hopping $t_{(00\frac12)}^{z z}=-0.684$~eV between the $z$-orbitals is the strongest, which is similar to that in La327 \cite{t-1,t-5,t-6}.
On the other hand, the crystal field splitting $\varepsilon_x-\varepsilon_z$ for both inner ($0.057$eV) and outer ($0.179$eV) layers are smaller than that in La327 ($0.367$eV) \cite{t-1,t-5,t-6,327arxivhu2}, indicating the two orbitals are more degenerate in La4310.

With these tight-binding parameters, we plot the band structure in Fig.~\ref{fig:tb}(a) together with the DFT band at $k_z=0$ for comparison.
The Fermi level is tuned to achieve the average filling $\av{n}=1.33$ per Ni-atom. Note that the tight-binding Fermi level is slightly higher than DFT since the $z$-directional dispersion has been neglected.
The Fermi surface of the tight-binding model is plotted in Fig.~\ref{fig:tb}(b), exhibiting four Fermi pockets. The electron-like $\alpha$-pocket around $\Gamma$ and hole-like $\gamma$-pocket around $M$ are mainly contributed by $x$ and $z$ orbitals, respectively, similarly to the case in La327. On the other hand, the single $\beta$ pocket in La327, contributed by both orbitals, now splits into two, which we label as $\beta$ and $\beta'$, respectively.
Due to the mirror symmetry with respect to the inner-plane, we can use its parity $P_z$ to label the different bands and Fermi pockets.
As shown in Fig.~\ref{fig:tb}(b), all of $\alpha$, $\beta$ and $\gamma$ pockets are $P_z$-even, and the $\beta'$ pocket is $P_z$-odd.
For the latter, the Bloch wave function by symmetry is $(1,0,-1)$ in the (top, middle, bottom) layer basis and hence is only contributed by outer-layers.
This is a marked feature not shared by La327.

{\it Electronic correlation effect}.
We now investigate the electronic correlation effect from the atomic multi-orbital Coulomb interactions
\begin{align}
H_I=&\sum_{i,a<b,\sigma\sigma'}\left( U'n_{ia\sigma}n_{ib\sigma'}+J_Hc_{ia\sigma}^\dag c_{ib\sigma} c_{ib\sigma'}^\dag c_{ia\sigma'} \right)\nonumber\\
&+\sum_{ia} Un_{ia\up}n_{ia\dn}+\sum_{i,a\ne b} J_Pc_{ia\up}^\dag c_{ia\dn}^\dag c_{ib\dn} c_{ib\up},
\end{align}
where $U$ is the intra-orbital Hubbard repulsion, $U'$ is the inter-orbital Coulomb interaction, $J_H$ is the Hund's coupling, and $J_P$ is the pair hopping interaction. They are assumed to respect the Kanamori relations $U=U'+2J_H$ and $J_H=J_P$ \cite{KanamoriRelations}. We use the constrained random phase approximation to estimate the characteristic value of $U$ to be in the range of 2eV to 4eV, so that the correlation is moderate. We use the SM-FRG to study the effective four-point one-particle irreducible vertices $\Gamma$ flowing against the running energy scale $\Lambda$. The Technical details can be found in Refs.~\cite{Wang_PRB_2012,Wang_PRB_2013,Tang_PRB_2019,t-2}.
At each $\Lambda$, the effective interactions $V_{\rm eff}$ in spin-density-wave (SDW), charge-density-wave (CDW) and superconductivity (SC) channels are extracted from the four-point vertices $\Gamma$.
The leading singular values $S$ of these effective interactions (as scattering matrices in the fermion bilinear basis) are monitored during the RG flow. The first divergence of $S$ out of all channels indicates the instability toward an electronic order characterized by the singular bilinear scattering mode, and the divergence energy scale is characteristic of the transition temperature $T_c$.

\begin{figure}
\includegraphics[width=\columnwidth]{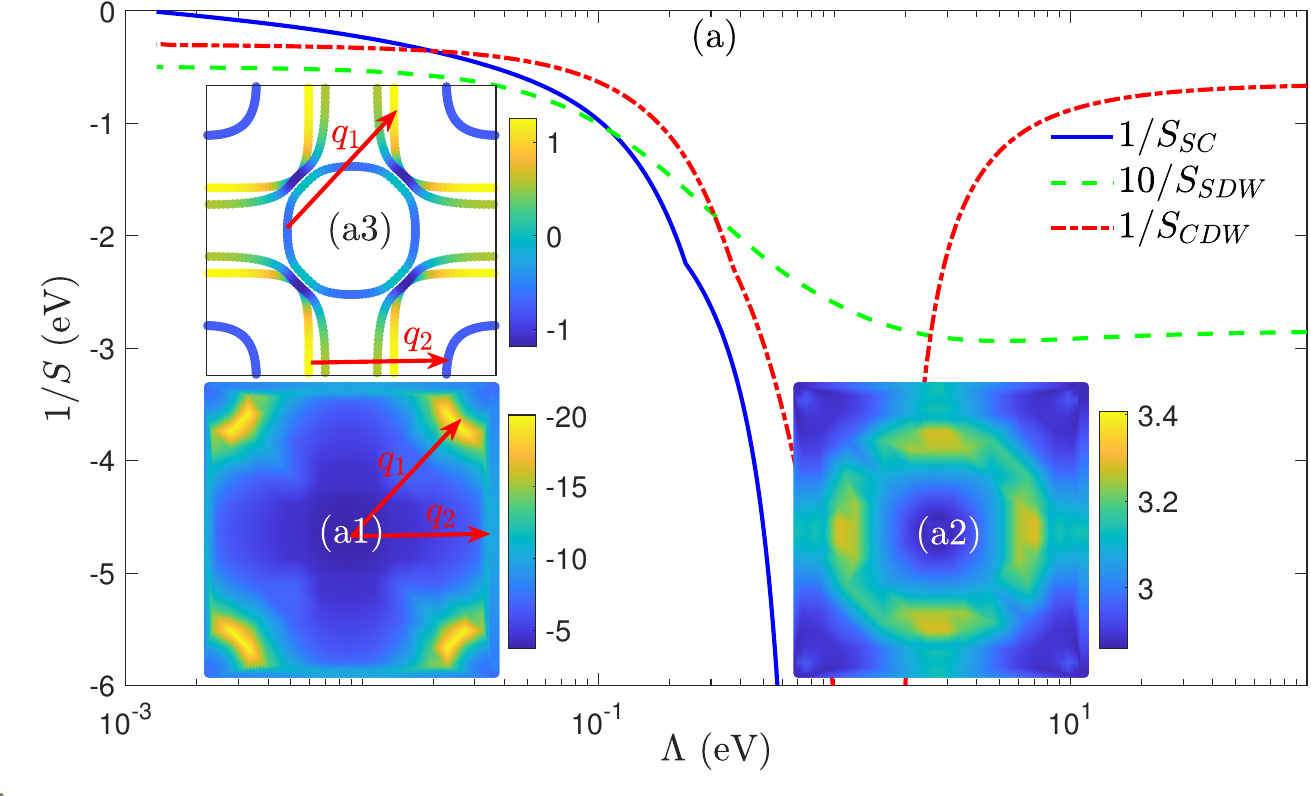}\\
\,\,\,\,\,\includegraphics[width=0.97\columnwidth]{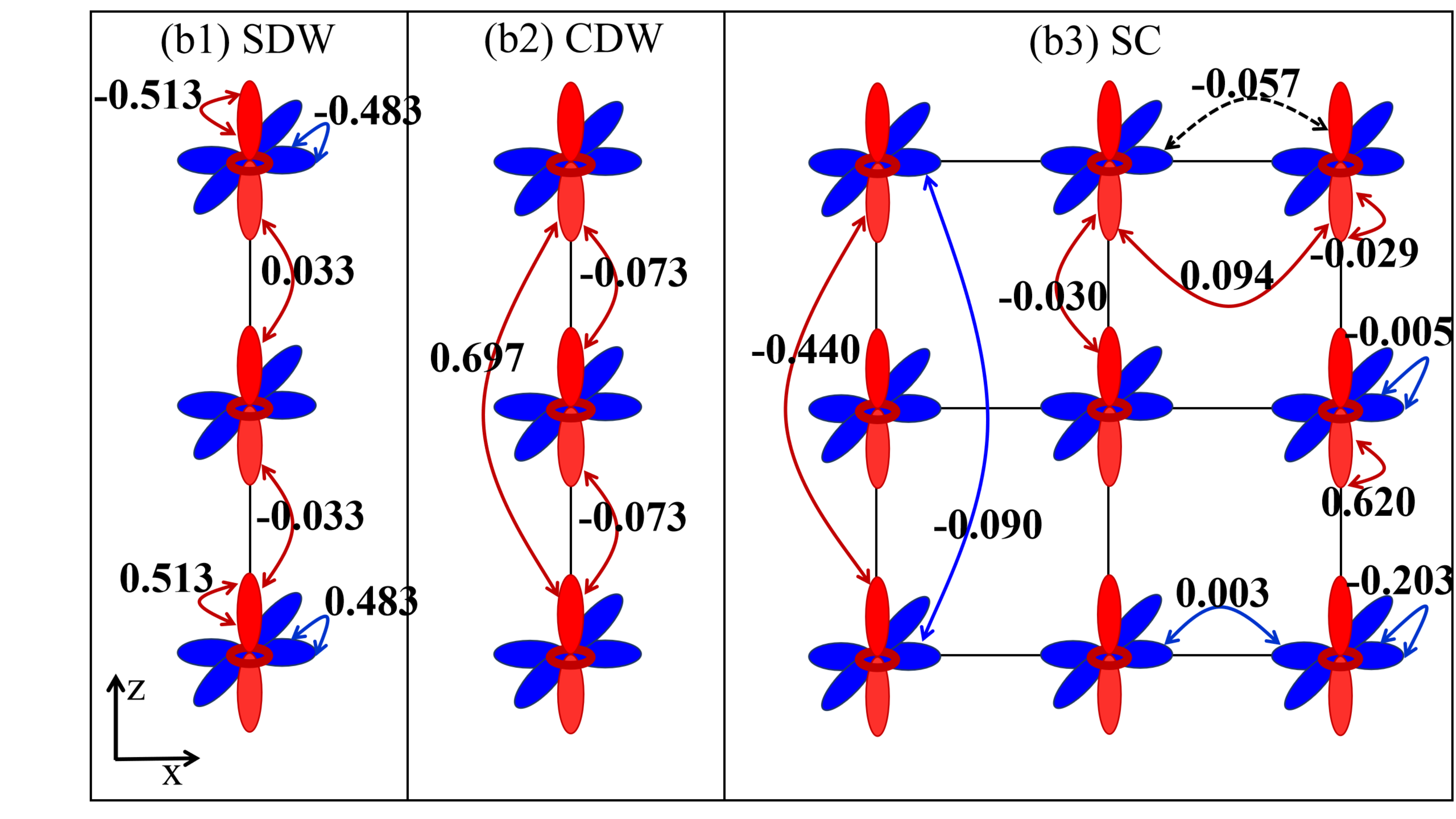}
\caption{(a) FRG flows of the inverse of the leading negative $S$ versus $\Lambda$ in the three channels, respectively. The $\0q$-dependence of $S$ in the SDW and CDW channels are plotted in subfigures (a1) and (a2), respectively. The SC gap function on the Fermi surface is plotted in the subfigure (a3). The corresponding patterns of the leading modes in the SDW, CDW and SC channels are sketched in (b1)-(b3), respectively.
}
\label{fig:frg}
\end{figure}

In Fig.~\ref{fig:frg}(a), we plot the inverse of the leading negative singular values $S$ versus $\Lambda$ in the three channels, respectively, at $U=3$~eV and $J_H=0.5$~eV.
At high energy scales, the SDW channel dominates.
As $\Lambda$ decreases, it grows up but saturates at low energy scales due to the lack of perfect Fermi surface nesting.
Instead, the SC channel is too weak and out of the view field at high energy scales, as the bare Coulomb interaction is repulsive. But as $\Lambda$ decreases to order of the band width, it is enhanced quickly as the SDW channel rises. Eventually, $1/S$ in the SC channel continues to grow logarithmically ally even if the other channels tend to saturate, a result of the standard Cooper mechanism. In the above sense, one could say the SC is induced by SDW fluctuations, as found also in La327.

In Fig.~\ref{fig:frg}(a1), we plot the momentum $\0q$-dependence of the leading negative singular value $S(\0q)$ in the SDW channel. Up to symmetry, there are two peaks at $q_1$ and $q_2$. 
Interestingly, we find the leading scattering modes at $q_1$ and $q_2$ are dominated by antiferromagnetic arrangement of outer-plane spins, in addition to the in-plane correlations, as sketched in Fig.~\ref{fig:frg}(b1).
Such antiferromagnetic correlations exist even in the bare spin susceptibility, and can be significantly enhanced by the Coulomb interactions due to the fact that the outer layers have less vertical hopping paths.
Note the spin structure is very different to the near-plane antiferromagnetic spin coupling in La327.

From Fig.~\ref{fig:frg}(a), the CDW channel is initially reduced at high energy scales because of the screening of charge density-density interactions. At lower energy scales, it rises again, and we find it corresponds to scattering of nonlocal fermion bilinears.
In particular, at the final stage of the RG flow,
the almost featureless $\0q$-dependence of $S(\0q)$ is shown in Fig.~\ref{fig:frg}(a2), and
the bilinear is mainly composed of $z$-orbitals on the outer planes as schematically plotted in Fig.~\ref{fig:frg}(b2).
This is a valence bond correlation consistent with the spin coupling between the outer planes discussed above.

In the SC channel, and from the singular scattering mode, we find the pairing is spin-singlet and belongs to the $A_{1g}$ representation of the $D_{4h}$ group. The pairing operator can be written explicitly as
\begin{align}
H_\Delta=&\sum_{amn\0k}\left[\Delta_{0,mn}^{aa} + \Delta_{1,mn}^{aa}f_{1+}(\0k)\right] c_{ma,\0k\up}^\dag c_{na,-\0k\dn}^\dag \nn\\
& \quad+\Delta_{1,mn}^{a\bar{a}}f_{1-}(\0k) c_{ma,\0k\up}^\dag c_{n\bar{a},-\0k\dn}^\dag + H.c. ,
\end{align}
where $m$/$n$ denotes layer index ($1$ to $3$ from top to bottom), $a$ denotes the $x$/$z$ orbital, $\bar{x}=z$ and $\bar{z}=x$,  $f_{1,\pm}(\0k)=2\cos k_x\pm 2\cos k_y$.
For $U=3$~eV and $J_H=0.5$~eV, we find the coefficients  (up to a global scale) $\Delta_{0,22}^{zz}=0.620$,
$\Delta_{0,13}^{zz}=\Delta_{0,31}^{zz}=-0.440$,
$\Delta_{0,11}^{xx}=\Delta_{0,33}^{xx}=-0.203$,
$\Delta_{0,11}^{zz}=\Delta_{0,33}^{zz}=-0.029$,
$\Delta_{1,11}^{zz}=\Delta_{1,33}^{zz}=0.094$,
$\Delta_{1,11}^{xx}=\Delta_{1,33}^{xx}=0.003$,
$\Delta_{1,11}^{xz}=\Delta_{1,11}^{zx}=\Delta_{1,33}^{xz}=\Delta_{1,33}^{zx}=-0.057$,
$\Delta_{0,12}=\Delta_{0,21}=\Delta_{0,32}=\Delta_{0,23}=-0.030$, etc.
The pairing pattern is sketched in Fig.~\ref{fig:frg}(b3).
For the $A_{1g}$ pairing symmetry, since $z$ and $x$ orbitals carry different representations ($A_{1g}$ and $B_{1g}$), only intra-orbital pairings are allowed for intra-unit-cell ones $\Delta_{0,mn}^{aa}$, which turn out to be much stronger than inter-unit-cell ones $\Delta_{1,mn}^{ab}$.
In particular, the strongest intra-unit-cell components are $\Delta_{0,22}^{zz}$ between $z$ orbitals in the inner-layer, and the vertical inter-layer pairing $\Delta_{0,13}^{zz}=\Delta_{0,31}^{zz}$ between two outer-layers as a result of the dominant antiferromagnetic coupling between top and bottom layers as shown above.
We further project the pairing function onto the band basis to obtain the gap function $\Delta_{n\0k}=\langle n,\0k| \sum_{\delta}\Delta_\delta e^{i\0k\cdot\bm{\delta}} |n,\0k\rangle$ where $|n,\0k\rangle$ is the Bloch state for band $n$, $\Delta_\delta$ is understood as a matrix in the orbital-layer basis, and we have used time-reversal symmetry $T$ for $|n,-\0k\rangle = T|n\0k\rangle$ to eliminate the numerical phase ambiguity.
The gap function on the Fermi surfaces is shown (with color scale) in Fig.~\ref{fig:frg}(a3).
The symmetry is evidently extended s-wave or $s_\pm$-wave, similar to that in La327.
Both $\alpha$- and $\gamma$-pockets are fully gapped with the same sign. The gap function on $\beta$ and $\beta'$-pockets are largely of opposite sign, although within a pocket it changes sign near the nodal direction. The SDW vectors (arrows) turn out to connect opposite gap signs on the Fermi surfaces, consistent with the spin-fluctuation scenario for singlet pairing.
The gap amplitude and gap sign change on the Fermi pockets can be checked in future experiments.

\begin{figure}
\includegraphics[width=\linewidth]{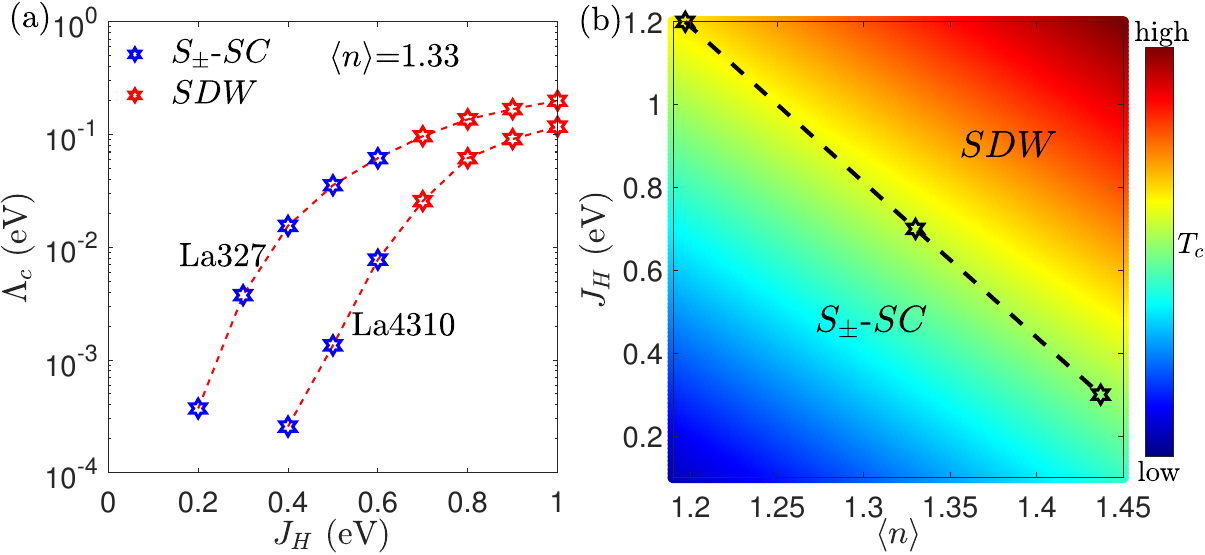}
\caption{ (a) Phase diagram by scanning $J_H$ for both La4310 and La327. (b) Phase diagram of La4310 with respect to the filling level $\av{n}$ and $J_H$, with $T_c$ qualitatively encoded by colors. In these calculations, we have used $U=3$~eV.}
\label{fig:phase}
\end{figure}

{\it Phase diagram}.
We have scanned the Hund's coupling $J_H$ by fixing $U=3$~eV to obtain the phase diagram as shown in Fig.~\ref{fig:phase}(a). The results of La327 are also presented for comparison. In both systems, larger $J_H$ enhances the SC until it drives the systems into the SDW state.
Interestingly, for the same interaction parameters, we find the $T_c$ of La4310 is always smaller than La327, in agreement with the experimental observation \cite{2023Nature,exp-32}.
Such a reduction of $T_c$ in La4310 can be understood by its longer-range, hence, weaker antiferromagnetic exchange between the top and bottom layers.
In this regard, increasing the layer number from La327 to La4310 does not enhance $T_c$.
This is in sharp contrast to trilayer cuprates, which are known to exhibit the highest $T_c$ among all cuprates \cite{ld-1,ld-2,ld-3,ld-4,ld-5}.
On the other hand, the single-layer La$_2$NiO$_4$ (La214) is know to be insulating. Therefore, we suggest further efforts to obtain/stabilize the bilayer structure phase to search for higher temperature superconductivity in nickelates.

We finally investigate the doping effect. The phase diagram with respect to the average filling level $\av{n}$ and Hund's coupling $J_H$ is plotted in Fig.~\ref{fig:phase}(b), with color-scaled $T_c$ qualitatively shown.
For all these fillings, SC is found to increase with $J_H$, and a large enough $J_H$ drives the system into the SDW state.
In addition, electron doping is also found to enhance $T_c$.
Actually, a large electron filling can produce an additional electron-like Fermi pocket around $\Gamma$, hence, enhancing the SDW fluctuations and increasing $T_c$ as a result. This doping dependence can be checked in future experiments.

{\it Conclusion}.
In summary, we have constructed a minimal tight-binding model based on the nickel $3d_{3z^2-r^2}$ and $3d_{x^2-y^2}$ orbitals and investigated the electronic correlation effect using functional renormalization group for La4310 under high pressure. The spin fluctuations are found to exhibit a peculiar antiferromagnetic coupling between the two outer-layers, and induce the $s_{\pm}$-wave superconductivity with the dominant pairings between $3d_{3z^2-r^2}$ orbitals both in the inner-plane and between the two outer-planes. The calculated $T_c$ in La4310 is systematically lower than in La327, suggesting La327 as the most promising candidate for higher $T_c$ in the various families of nicklates.

\begin{acknowledgments}
This work is supported by National Key R\&D Program of China (Grant No. 2022YFA1403201), National Natural Science Foundation of China (Grant No. 12374147, No. 12274205, No. 12074213, No. 92365203, No. 11874205, and No. 11574108), and Major Basic Program of Natural Science Foundation of Shandong Province (Grant No. ZR2021ZD01).
\end{acknowledgments}



\bibliography{LaNiO}

\end{document}